\documentclass{jaa}

\usepackage{graphicx}
\usepackage{xspace}
\usepackage{epsf,url,wrapfig ,fancyhdr,multirow,lscape,appendix,rotating, setspace,amsmath,amssymb}
\usepackage{graphics,epsfig,url}
\usepackage{graphicx}
\usepackage{gensymb}
\usepackage{amssymb}
\usepackage{xcolor}
\usepackage{color}
\usepackage{lipsum}
\usepackage{textcomp}

\begin{document}\sloppy


\title{Evidence of Heavy Obscuration in the Low-Luminosity AGN NGC~4941}

\author{Arghajit Jana\textsuperscript{1}, Sachindra Naik\textsuperscript{1}, and Neeraj Kumari\textsuperscript{1,2}}
\affilOne{\textsuperscript{1}Astronomy and Astrophysics Division, Physical Research Laboratory, Navrangpura, Ahmedabad - 380009, India\\}
\affilTwo{\textsuperscript{2}Indian Institute of Technology, Gandhinagar, Gujarat, India\\}

\twocolumn[{

\maketitle

\corres{argha0004@gmail.com}

\msinfo{***}{***}

\begin{abstract}
We present the results obtained from timing and spectral studies of the highly obscured low luminosity active galactic nucleus NGC~4941 using data obtained from the {\it Nuclear Spectroscopic Telescope Array} and the {\it Neil Gehrels Swift} Observatories. We find similar variability in $3-10$~keV and $10-60$~keV energy ranges with fractional rms variability of $\sim$14\%. We investigate broad-band spectral properties of the source in 3-150 keV range, using data from {\it NuSTAR} and {\it Swift}/BAT, with phenomenological slab model and physically motivated {\tt mytorus} model. From the spectral analysis, we find heavy obscuration with global average column density of the obscured material as $3.09^{+1.68}_{-1.01} \times 10^{24}$ cm$^{-2}$. Evidence of a strong reflection component is observed in the spectrum. We detect a strong iron line with equivalent width of $\sim$1~keV. From the slab model, we obtain the exponential cutoff energy as $177^{+92}_{-16}$~keV. From this, we estimate the Compton cloud properties with the hot electron temperature $kT_{\rm e} = 59^{+31}_{-5}$~keV and the optical depth $\tau = 2.7^{+0.2}_{-1.6}$.

\end{abstract}


}]


\doinum{12.3456/s78910-011-012-3}
\artcitid{\#\#\#\#}
\volnum{000}
\year{0000}
\pgrange{1--}
\setcounter{page}{1}
\lp{1}

\section{Introduction}

Active galactic nuclei (AGNs) are believed to be powered by accretion onto the supermassive black holes (Rees 1984) that reside at the center of the galaxies. The accretion disk predominately emits in the UV/optical wavebands (Shakura \& Sunyaev 1973). The UV/optical seed photons from the accretion disk undergo inverse Comptonization in the Compton corona producing emission in X-rays (Sunyaev \& Titarchuk 1980, Haardt \& Maraschi 1993). The emitted X-ray photons can be well approximated with a power-law continuum. A fraction of the emitted X-ray photons is reflected in the surrounding materials, producing Fe K-line complex in $\sim 6-8$~keV range and a reflection hump in $\sim 15-40$~keV range (George \& Fabian 1991, Matt et al. 1991). Additionally, an excess in the soft X-ray ($<1$~keV), known as soft-excess, is also observed (Singh et al. 1985).

Based on the optical observations, the AGNs are broadly classified into two classes depending on the presence or absence of broad emission lines. The broad emission lines (originate in the broad line emitting regions or BLRs) are observed in type-1 AGNs, while it is absent in the type-2 AGNs. The unified model (UM) of AGNs explains different classes of AGNs based on the orientation of viewing angle (Antonucci et al. 1985, 1993). In this model, a dusty torus surrounds the nuclear region at a parsec scale. The type-1 AGNs are observed face-on where the BLRs are visible, while the type-2 AGNs are viewed edge-on where the torus obscures the BLR (Awaki et al. 1991). Additionally, the narrow emission lines (originate in narrow line emitting regions or NLRs) are observed in both types of AGNs.

In the X-ray wavebands, the classification of AGNs (obscured or un-obscured) is based on the value of equivalent hydrogen column density ($N_{\rm H}$) along the line of sight which characterizes the obscured material surrounding the AGNs. If the column density $N_{\rm H} < 10^{22}$ cm$^{-2}$, the AGN is an un-obscured AGN, whereas if $N_{\rm H} > 10^{22}$ cm$^{-2}$, the AGN is classified as an obscured AGN. The obscured AGNs can be further classified as Compton-thin or Compton-thick (CT). If the column density $N_{\rm H} > 10^{24}$ cm$^{-2}$, the AGN is classified as Compton-thick AGN (e.g., Hickox et al. 2018).

Recently, a new sub-class of AGNs has emerged, known as changing-look AGNs (CLAGNs). In optical changing-look events (hereafter, changing-state event; CS event), the type-1 (or type 1.2/1.5) AGNs transit to type-2 (or type 1.8/1.9) and vice versa with the disappearance or appearance of the broad emission lines. Several nearby galaxies, such as NGC~1566 (Parker et al. 2019), NGC~3516 (Ilic et al. 2020), Mrk~590 (Denney et al. 2014), NGC~2617 (Shappee et al. 2014), Mrk~1018 (Noda et al. 2018) have been found to show such peculiar behavior. A different type of changing-look events have been observed in the X-ray wavebands, with an AGN switching between Compton-thin ($N_{\rm H}< 10^{24}$ cm$^{-2}$) and Compton-thick (CT; $N_{\rm H}> 10^{24}$ cm$^{-2}$) states (Risaliti et al. 2002; Matt et al. 2003). The X-ray CL events (hereafter, changing-obscuration event; CO event) have been observed in many AGNs, namely IC~751 (Ricci et al. 2016), NGC~4507 (Braito 2013), NGC~6300 (Jana et al. 2020).

Low luminosity AGNs (LLAGNs; bolometric luminosity $L_{\rm bol} < 10^{43}$ erg s$^{-1}$) holds key to understand the changing-state events. Many changing-state AGNs are observed to remain in the low-luminosity state over decades before showing a CS event. Recently, NGC~1566 showed an outburst in June 2018 after remaining in the low luminosity state for over a decade (Jana et al. 2021). During the outburst, the optical, UV, and X-ray flux increased by $\sim 25-30$ times compared to the low luminosity state with the reappearance of broad emission lines (Oknyansky et al. 2020). Similar behavior is also observed in NGC~3516 (Ilic et al. 2020). Thus, it is essential to study the accretion properties of low-luminosity AGNs.

NGC~4941 is classified as a Seyfert~2 galaxy (Veron-cetty et al. 2006) with a morphological class of Sa (Fisher et al. 2008). Over the past years, NGC~4941 was observed with several X-ray observatories, such as {\it ASCA} (Terashima et al. 2002), {\it BeppoSAX} (Maiolino et al. 1998), {\it Chandra} (Bottacini et al. 2012), {\it Suzaku} (Kawamuro et al. 2013) and {\it NuSTAR} (Georgantopoulos et al. 2019, Garcia-Burillo et al. 2021). {\it The ASCA}, {\it BeppoSAX}, {\it Suzaku} and {\it NuSTAR} observations revealed a heavily obscured nucleus with line of sight column density $N_{\rm H}>10^{23}$ cm$^{-2}$ and a strong Fe K$\alpha$ emission line in the spectrum. The mass of the black hole in NGC~4941 is reported to be $\sim 10^{6.9}~M_{\odot}$ (Asmus et al. 2011). While investigating the obscuration properties of {\it Swift}/BAT AGNs, several authors used {\it NuSTAR} observations of Swift/BAT AGNs, and carried out systematic spectral studies to understand the nature of the torus of the AGN sample (e.g., Georgantopoulos et al. 2019, Balokovic et al. 2020, Garcia-Burillo et al. 2021, Zhao et al. 2021). Though the {\it NuSTAR} observation of NGC~4941 on 19 January 2016 was used in above studies, a detailed investigation of the source properties was not carried out. Considering this, we performed spectral and timing studies of NGC~4941 in broad energy range using data from {\it NuSTAR} (on 19 January 2016) and {\it Swift}/BAT observations.

The paper is organized in the following. In \S2, we describe the observation and data extraction process. We present the results obtained from the timing and spectral analysis in \S3. In \S4, we discuss our findings. The summary of the work is presented in \S5.

\section{Observation and Data Extraction}

NGC~4941 was observed with {\it NuSTAR} on 19 January 2016 for an exposure of $\sim 21$~ks. {\it NuSTAR} is a hard X-ray focusing telescope consisting of two identical modules: FPMA and FPMB (Harrison et al. 2013). We reprocessed the raw data with the {\it NuSTAR} Data Analysis Software ({\tt NuSTARDAS}, version 1.4.1). Cleaned event files were generated and calibrated using the standard filtering criteria in the {\tt nupipeline} task and the latest calibration data files available in the {\it NuSTAR} calibration database (CALDB)\footnote{http://heasarc.gsfc.nasa.gov/FTP/caldb/data/nustar/fpm/}. We extracted the source and background products considering circular regions with radii 60 arcsec and 90 arcsec, respectively. The spectra and light curves were extracted using the {\tt nuproduct} task. The light curves were binned at 500~s. To improve the S/N, we co-added FPMA and FPMB spectra using {\tt ADDASCASPEC} task. We re-binned the spectra with 20 counts per bin by using the {\tt grppha} task. We analysed {\it NuSTAR} spectrum in $3-60$~keV energy range. We also included the spectrum of the source from {\it Swift}/BAT in $14-150$~keV range, obtained from 105-month BAT survey (Oh et al. 2018)\footnote{https://swift.gsfc.nasa.gov/results/bs105mon/653}.

\section{Result}
We studied NGC~4941 using the data obtained from the {\it NuSTAR} and {\it Swift}/BAT in a combined energy range of $3-150$~keV. In our study, we used following Cosmological parameters: $H_{\rm 0} = 70$ km s$^{-1}$ Mpc$^{-1}$, $\Lambda_{\rm 0}=0.73$ and $\Omega_{\rm M}=0.27$ (Bennett et al. 2003).

\subsection{Timing Properties}
\begin{figure}
\centering
\includegraphics[width=8.5cm]{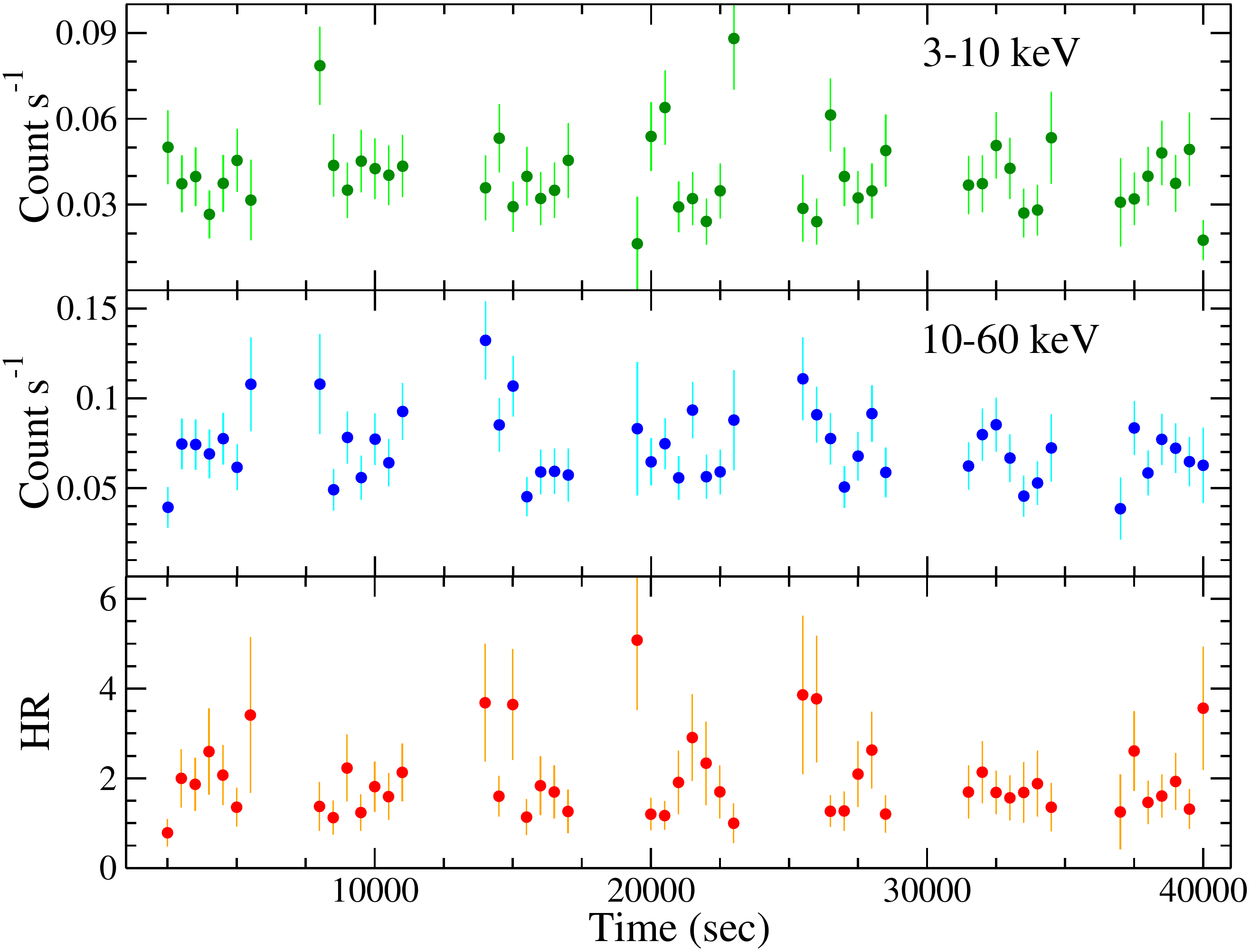}
\caption{Light curves of NGC~4941 in $3-10$~keV (top panel) and $10-60$~keV (bottom panel) ranges, obtained from {\it NuSTAR} observation, are shown. Bottom panel shows the variation of the hardness ratio.}
\label{lc}
\end{figure}

To study time variability in the source, we generated light curves in $3-10$~keV and $10-60$~keV energy ranges with a time-resolution of 500~s from {\it NuSTAR} data. Figure~1 shows the light curves in $3-10$~keV and $10-60$~keV energy ranges in the top and middle panels, respectively. In the bottom panel of Figure~\ref{lc}, we show the variation of hardness ratio (HR) which is defined as the ratio between the count rates in $10-60$~keV and $3-10$~keV bands. To study the variability, we calculated the fractional rms amplitude ($F_{\rm var}$) for the light curves in different energy ranges (Vaughan et al. 2003). Both energy bands showed similar variabilities with $F_{\rm var} = 14.4 \pm 4.8 \%$ and $14.5 \pm 7.4 \%$ in $3-10$~keV and $10-60$~keV ranges, respectively. 

\subsection{Spectral Properties}

We carried out spectral analysis of NGC~4941 using HEASEARC's spectral analysis package {\tt XSPEC}v12.10\footnote{https://heasarc.gsfc.nasa.gov/xanadu/xspec/} (Arnaud 1996). In our spectral analysis, we used data in $3-60$~keV and $14-150$~keV ranges from the {\it NuSTAR} and {\it Swift}/BAT observations, respectively, to get a broad-band coverage of $3-150$~keV energy range. All the errors are quoted at 1.6$\sigma$ level (90\% confidence).

\subsection*{Slab Model}
We started our analysis with phenomenological models (hereafter slab model) {\tt powerlaw} and {\tt pexrav} (Magdziarz \& Zdziarski 1998). The slab model consists of an absorbed power-law continuum with an exponential cut-off, reflection from cold materials and Fe K-line. Additionally, we also included a scattered emission which is often observed in the Compton thin sources (Turner et al. 1997). We did not include any component for the soft-excess as we study in the energy range of 3--150~keV. The cold reflection component was modeled with {\tt pexrav} component. {\tt pexrav} describes the reflection from a cold semi-infinite slab. The power-law photon index, cut-off energy and normalization were tied up with the primary continuum. We set the reflection fraction ($R_{\rm refl}$) to a negative value, thus, {\tt pexrav} only considered as the reflection component. We fixed the abundances to the Solar value and inclination angle to 60$^{\circ{}}$. Thus, the only free parameter in {\tt pexrav} was the reflection fraction ($R_{\rm refl}$). We modeled the scattered emission with a power-law with exponential cut-off and tied the parameters with the primary emission. The scattering fraction ($f_{\rm Scat}$) was estimated by using a multiplicative constant with the scattering emission. Our complete model can be written in {\tt XSPEC} as:

{\bf constant1 * phabs1 * (zphabs*cabs*zcutoff + pexrav + zgauss + constant2*zcutoff)}.

Here, constant1 and constant2 are the cross-normalization between {\it NuSTAR} \& BAT, and scattered fraction, respectively. {\tt phabs1} represents the Galactic absorption in the direction of the source which we fixed at the Galactic value of $2.25 \times 10^{20}$ cm$^{-2}$ (HI4PI Collaboration 2016). The {\tt zphabs} and {\tt cabs} components represent absorptions due to the line of sight column density and Compton scattering, respectively. The column densities of these two models were tied together. During our analysis, we fixed the line width of Gaussian component at 0.1~keV.

The spectral analysis with the slab model gave us a good fit with a $\chi^2=164$ for 161 degrees of freedom (dof). We obtained the photon index $\Gamma = 1.47 \pm 0.05$, cut-off energy, $E_{\rm cut} = 177^{+92}_{-16}$ and reflection fraction $R_{\rm refl} < 0.89$. The Fe K$_\alpha$ line was detected at $6.33 \pm 0.05$~keV with equivalent width EW=$0.71^{+0.03}_{-0.01}$~keV. We also found the source in the Compton-thin state with the line of sight column density $N_{\rm H} = (7.5 \pm 0.3) \times 10^{23}$ cm$^{-2}$. The best-fit parameters obtained from our spectral fitting are presented in Table~1. Figure~2 shows the best-fitted spectrum with slab model in the top left panel. Corresponding residuals are shown in the bottom left panel of the figure. The right panel of Figure~2 shows the unfolded spectrum. The black, blue, red, green and magenta lines represent the total, primary continuum, reprocessed emission, Fe K$\alpha$ line, and scattered emission, respectively. Figure~3 shows the confidence contour between the photon index ($\Gamma$) and cut-off energy ($E_{\rm cut}$).

\begin{figure*}
\centering
\includegraphics[width=8.5cm]{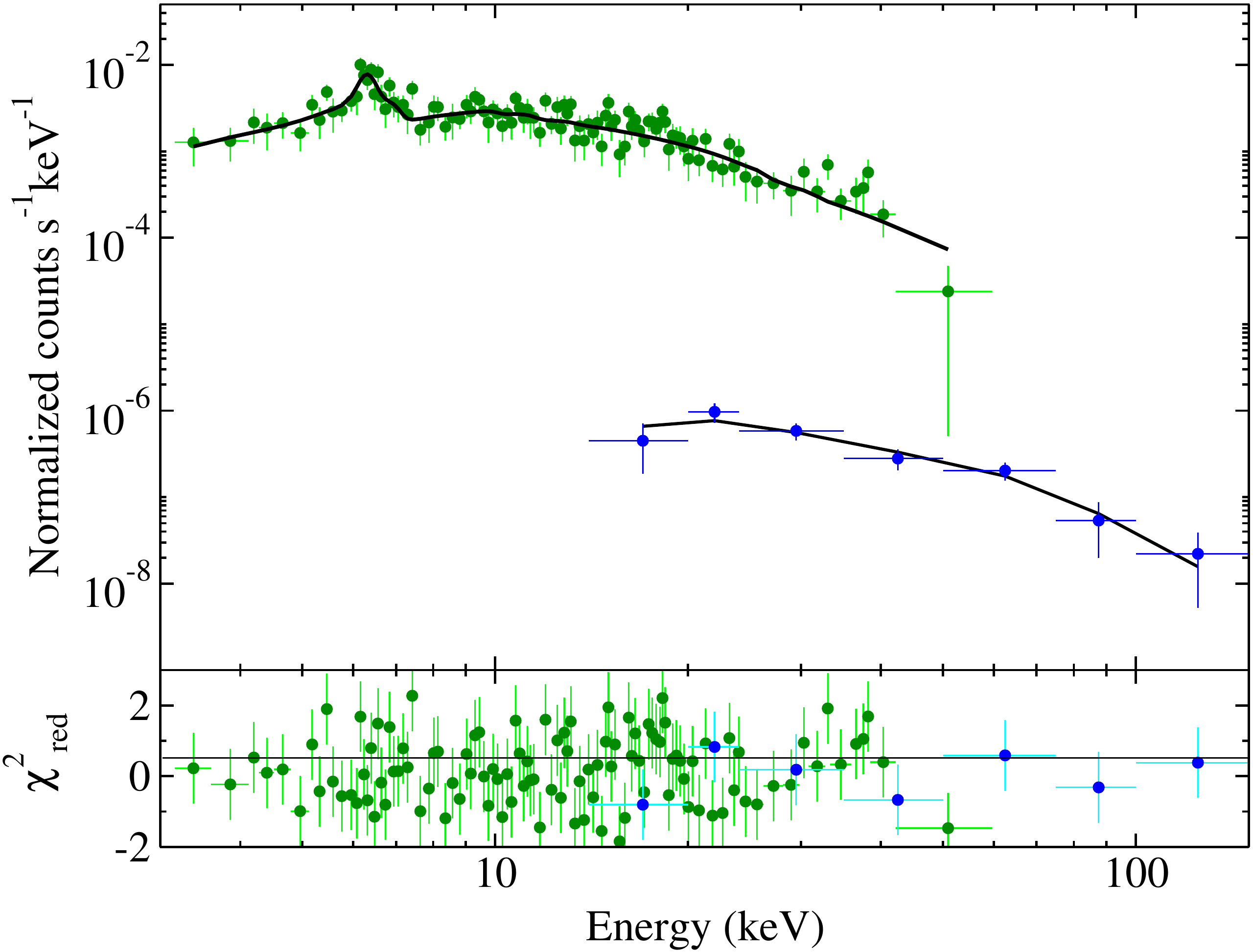}
\includegraphics[width=8.5cm]{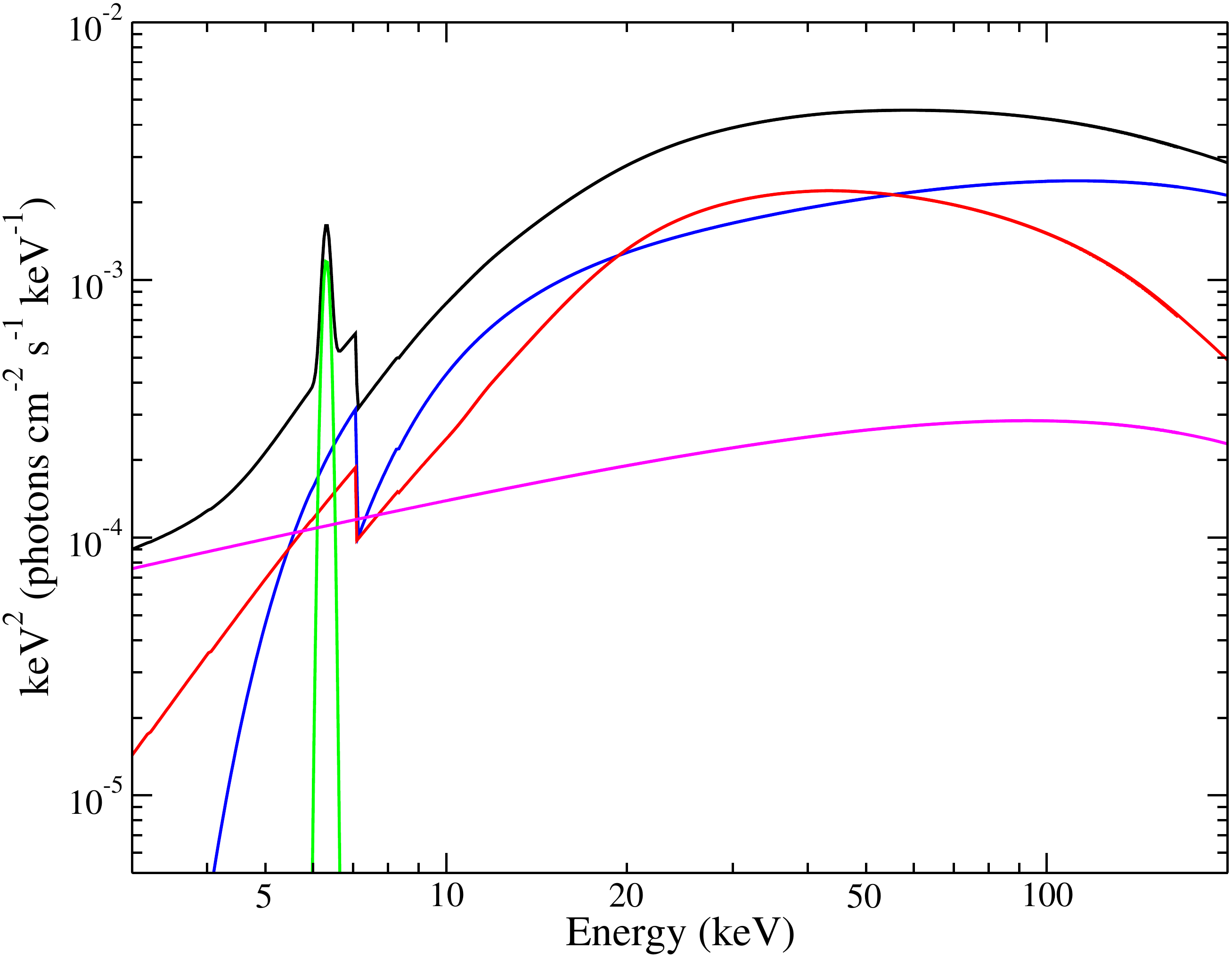}
\caption{Left side : Upper panel -- Slab model fitted spectrum (black solid line). Green and blue circles represent the {\it NuSTAR} and {\it Swift}/BAT data, respectively. Bottom panel -- Residuals in terms of data--model/error. Right side : The unfolded spectrum. The black, blue, red, green and magenta lines represent the total, primary continuum, reprocessed emission, Fe K$_\alpha$ line, and scattered emission, respectively.}
\end{figure*}

\begin{figure}
\centering
\includegraphics[angle=270,width=8.5cm]{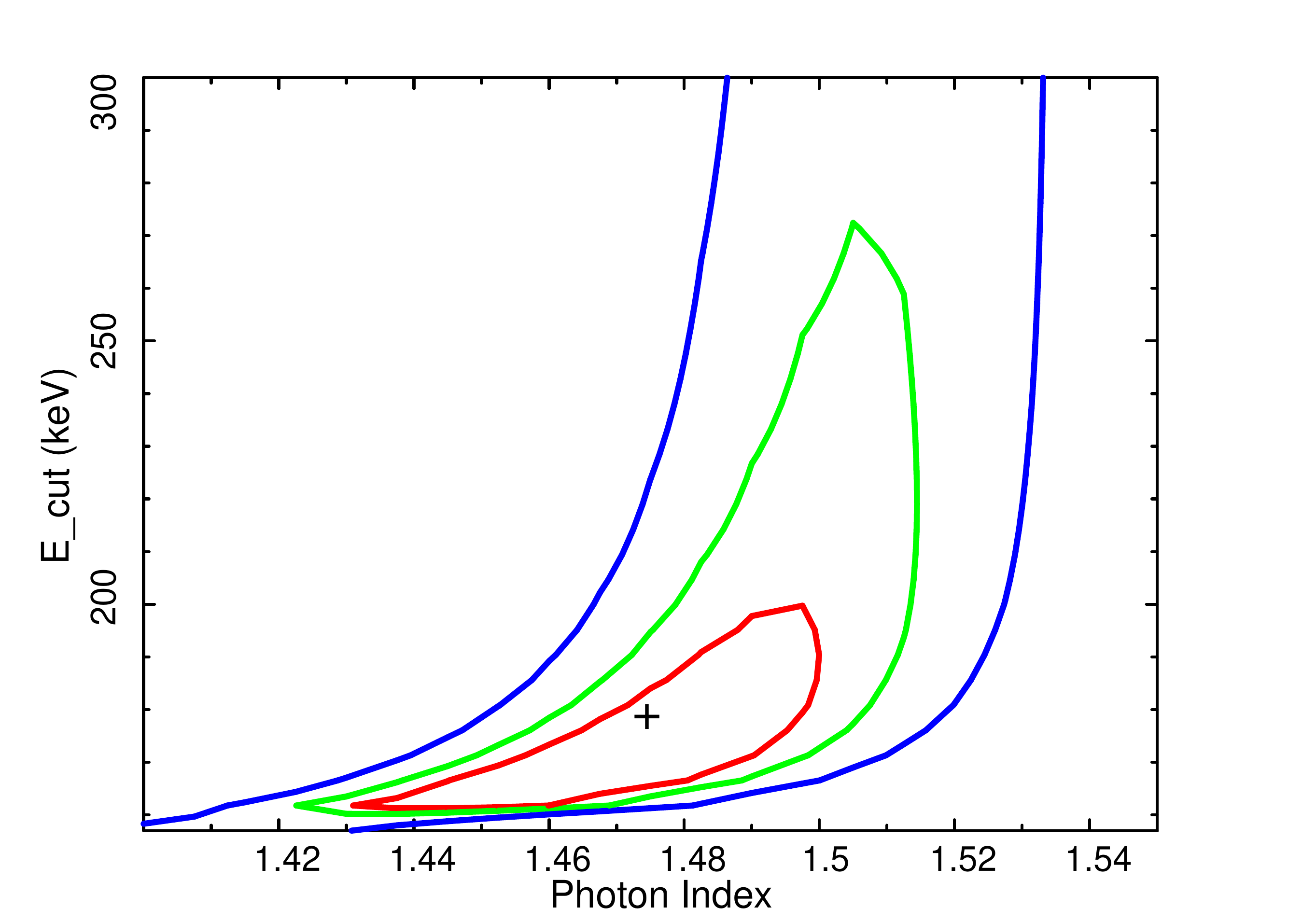}
\caption{Confidence contours between the photon index ($\Gamma$) and cut-off energy ($E_{\rm cut}$).}
\end{figure}

\begin{table}
\centering
\caption{Spectral Analysis Results}
\begin{tabular}{lll}
\hline
Parameters    &  Slab & MYTorus \\
\hline
$C_{\rm BAT}$   & $1.12\pm0.22$  & $1.11^{+0.34}_{-0.29}$ \\ 
$N_{\rm H}^{\rm los}$ ($10^{24}$ cm$^{-2}$) & $0.75^{+0.33}_{-0.29}$ & $0.76\pm0.09$ \\
$N_{\rm H}^{\rm tor}$ ($10^{24}$ cm$^{-2}$) & -- & $3.09^{+1.68}_{-1.01}$\\
Photon index $\Gamma$ & $1.47\pm0.05$& $1.71\pm0.10$\\
$E_{\rm cut}$ (keV)  & $177^{+92}_{-16}$ & --\\
$N_{\rm PL}$  & $6.08\pm0.12$& $6.73\pm0.15$ \\
 $R_{\rm refl}$ & $<0.89$ & --\\
$A_{\rm S}$ & -- & $3.15\pm0.32$ \\
$f_{\rm Scat}$ ($10^{-2}$) & $8.04^{+0.35}_{-0.78}$ & $9.32^{+0.35}_{-0.41}$ \\
Fe K$\alpha$ EW (keV) & $0.71^{+0.03}_{-0.01}$ & $1.03^{+0.07}_{-0.13}$ \\
$\chi^2$/dof & 165/161 & 165/162\\
\hline
$F_{\rm 2-10}^{\rm obs}$  & $7.52\pm0.08$ & $7.54\pm0.09$ \\
$L_{2-10}^{\rm int}$      & $2.12\pm0.61$ & $2.13\pm0.66$ \\
$L_{\rm bol}$             & $3.18\pm0.92$ & $3.19\pm0.99$ \\
\hline
\end{tabular}
\caption{All the errors are quoted at 90\% confidence (1.6$\sigma$) level. $C_{\rm BAT}$ is cross-normalization factor of BAT and NuSTAR. $N_{\rm H}^{\rm los}$ and $N_{\rm H}^{\rm tor}$ are the line of sight column density and averaged global column density of the torus, respectively. $N_{\rm PL}$ (Power-law Normalization) is in the unit of $10^{-4}$ ph cm$^{-2}$ s$^{-1}$). $F_{\rm 2-10}^{\rm obs}$ (observed flux in 2-10 keV range) is in unit of $10^{-13}$ ergs cm$^{-2}$ s$^{-1}$. $L_{2-10}^{\rm int}$ (intrinsic luminosity in 2-10 keV range) is in unit of $10^{41}$ ergs s$^{-1}$. $L_{\rm bol}$ (bolometric luminosity) is in unit of $10^{42}$ ergs s$^{-1}$.}
\label{tab:result}
\end{table}

\begin{figure*}
\centering
\includegraphics[width=8.5cm]{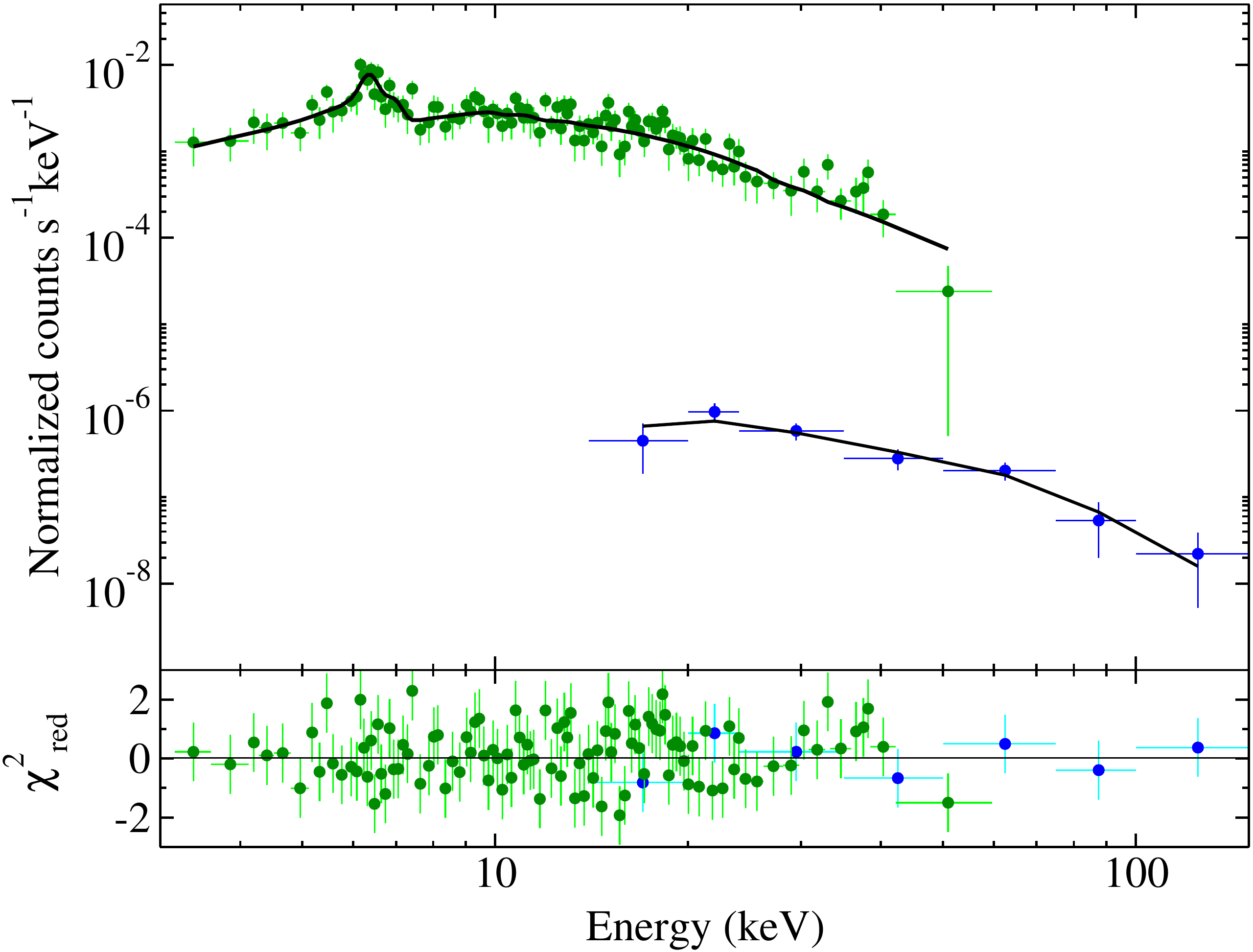}
\includegraphics[width=8.5cm]{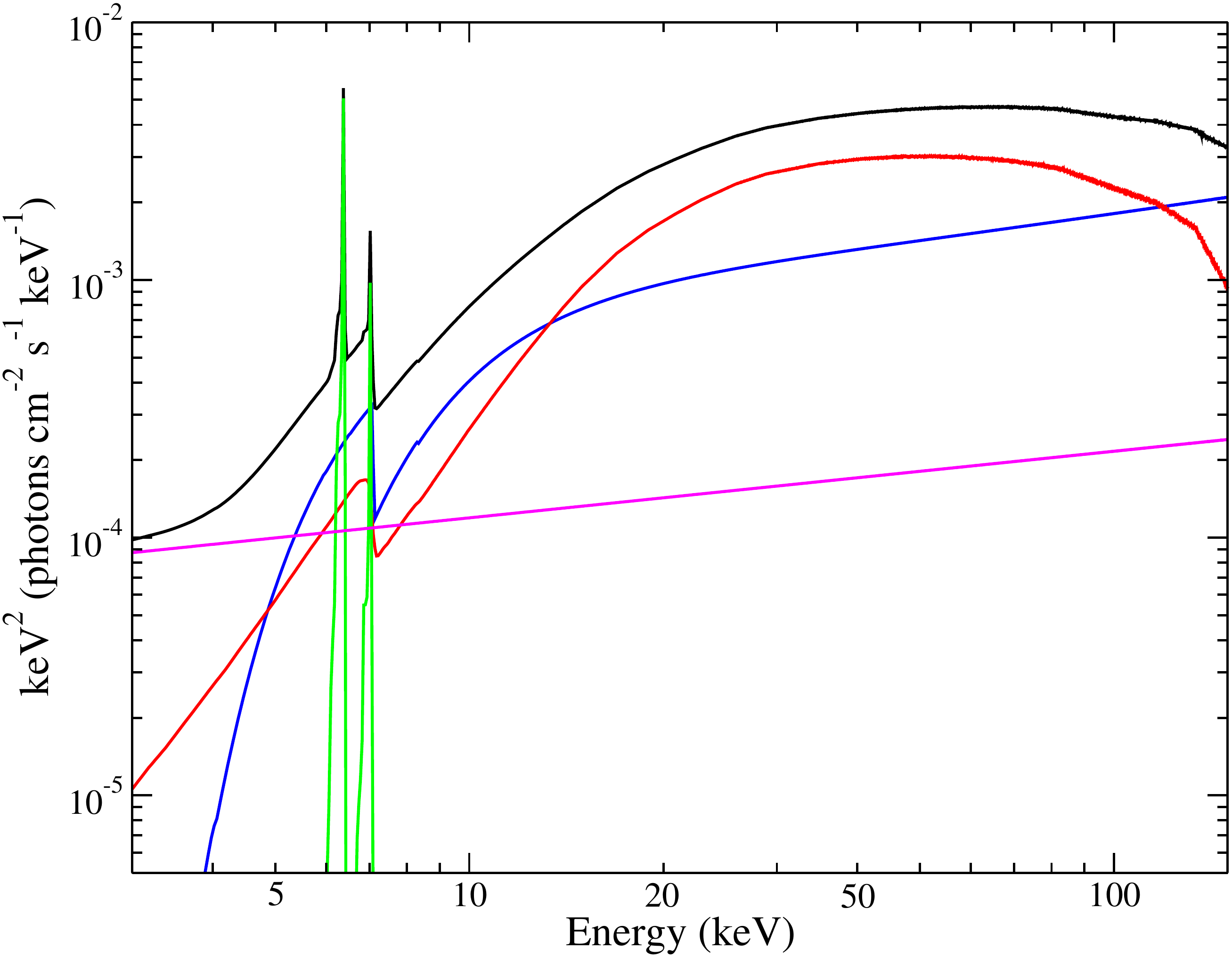}
\caption{Left side : Upped panel -- MYTorus model fitted spectrum (black solid line) is shown. Green and blue circles represent the {\it NuSTAR} and {\it Swift}/BAT data, respectively. Bottom panel -- Residuals in terms of data--model/error. Right side : The unfolded spectrum is shown. The black, blue, red, green and magenta lines represent the total, primary continuum, reprocessed emission, Fe K$_\alpha$ \& Fe K$_\beta$ lines, and scattered emission, respectively.}
\end{figure*}

\begin{figure}
\centering
\includegraphics[angle=270,width=8.5cm]{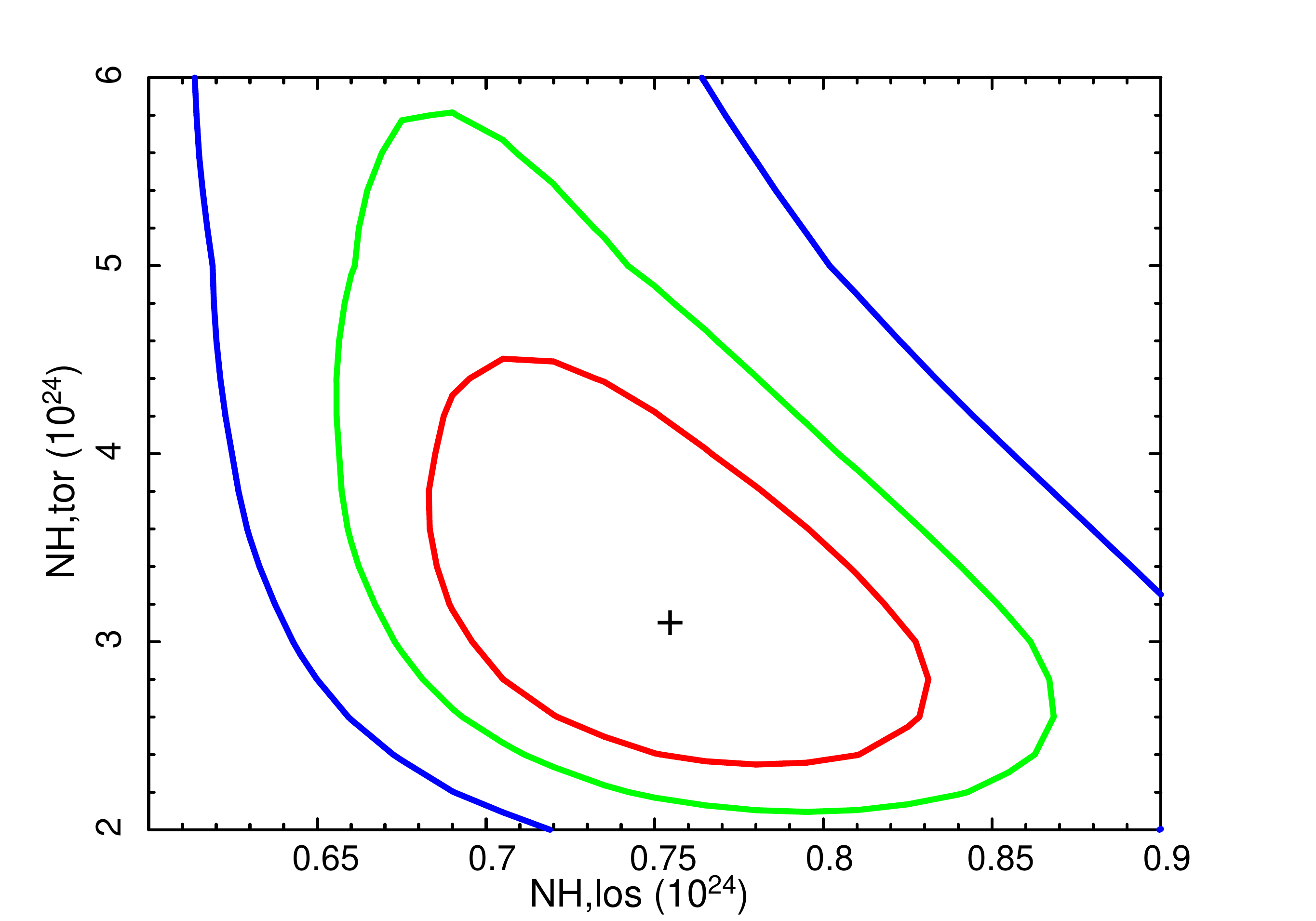}
\caption{Confidence contours between $N_{\rm H}^{\rm los}$ and $N_{\rm H}^{\rm tor}$.}
\end{figure}

\subsection*{Torus Model}
The {\tt pexrav} model does not consider the complex structure of the torus. Hence, the slab model does not give us complete information on the reprocessed emission or obscured materials. Various torus-based models have been developed considering the complex structure and geometry of the torus and estimated the reprocessed emission self-consistently (Murphy \& Yaqoob 2009; Brightman et al. 2011; Paltani \& Ricci 2017; Atsushi et al. 2019). Here, we studied NGC~4941 with physically motivated torus-based model {\tt mytorus} (Murphy \& Yaqoob 2009; Yaqoob 2012)\footnote{https://mytorus.com}.

The {\tt mytorus} model considers a dusty absorbing torus with uniform density surrounding the nucleus with a fixed covering factor 0.5 i.e., half-opening angle of 60$^{\circ{}}$. The {\tt mytorus} model has three components: absorbed primary emission or zeroth ordered component ({\tt mytz}), reprocessed emission ({\tt myts}), and an iron fluorescent line emission ({\tt mytl}; Fe K$_\alpha$ and Fe K$_\beta$). One can use the {\tt mytorus} model in two configurations: coupled and decoupled. The coupled configuration considers a uniform torus, while the decoupled configuration describes a clumpy or patchy torus. Many studies suggest that the torus is not uniform, rather made up of clumps of gas as opposed to what the UM had suggested (Krolik et al. 1988; Laor 1993; Honig et al. 2007, 2012; Nenkova et al. 2008ab). Thus, we applied the decoupled configuration of the {\tt mytorus} model to the combined spectra of NGC~4941. The model is represented as follows:

{\bf Constant1 * phabs1 * ( powerlaw*MYTZ + $A_{\rm S}$*MYTS + $A_{\rm L}$*gsmooth*MYTL + constant2*powerlaw )}.

Here, the photon index ($\Gamma$) and normalization of all four components are tied together. The inclination angle becomes a dummy parameter in this configuration (Yaqoob 2012). Hence, the inclination of the {\tt mytz} was set to 90$^{\circ{}}$, while the inclination of {\tt myts} and {\tt mytl} were set to 0$^{\circ{}}$. The column density of the {\tt mytz} component represents the line of sight column density ($N_{\rm H}^{\rm los}$). The column densities of the {\tt myts} and {\tt mytl} components were tied together. This column density represents the averaged global column density of the obscured materials ($N_{\rm H}^{\rm tor}$). The {\tt myts} component represents the reprocessed emission from the backside of the torus. As recommended, the relative normalization of the reflected ($A_{\rm S}$) and line emission ($A_{\rm L}$) components were tied, i.e. $A_{\rm S} = A_{\rm L}$. The last term represents the scattered primary emission as in the slab model.

The spectral fitting with this model gave us a good fit with $\chi^2 = 165$ for 162 dof. The line of sight column density and averaged column density of the obscured material were estimated to be $N_{\rm H}^{\rm los}= (0.76\pm0.09) \times 10^{24}$ cm$^{-2}$ and $N_{\rm H}^{\rm tor} = 3.09^{+1.68}_{-1.01} \times 10^{24}$ cm$^{-2}$, respectively. The photon index ($\Gamma = 1.71\pm0.10$) was found to be flatter compared to the slab model. The relative normalization of the reprocess emission was obtained to be $A_{\rm S} = 3.15\pm0.32$. 

As Fe K$_\alpha$ and Fe K$_\beta$ lines are calculated self-consistently in {\tt mytorus} model, the EW cannot be estimated in a straightforward way. The EW is calculated in the observed frame by taking the ratio of the total continuum flux to the line flux in 6.08--6.72~keV range (0.95 $E_{\rm K_\alpha}$ -- 1.05 $E_{\rm K\alpha}$). Then EW is calculated in the rest frame energy by multiplying with $(1+z)$. A strong iron line was also observed with {\tt mytorus} model with EW = $1.03^{+0.07}_{-0.13}$~keV. The FWHM of Fe K$\alpha$ line was obtained from the convolution model {\tt gsmooth}, with FWHM = 117.7 $\sigma_{\rm L}$(eV) km s$^{-1}$. We found the FWHM of the line as $<4900$ km s$^{-1}$. The results obtained from a detailed spectral analysis of the decoupled configuration of {\tt mytorus} model is presented in Table~1. The best-fitted spectrum with {\tt mytorus} model and corresponding residuals are shown in the top and bottom panels of the left side of Figure~4. The panel in the right side of Figure~4 shows the unfolded spectrum, fitted with the decoupled configuration of {\tt mytorus} model. The black, blue, red, green and magenta lines in the panel represent the total, primary continuum, reprocessed emission, Fe K$_\alpha$ \& Fe K$_\beta$ lines, and scattered emission, respectively. Figure~5 shows the confidence contours between the line of sight column density ($N_{\rm H}^{\rm los}$) and averaged global column density of the obscured material ($N_{\rm H}^{\rm tor}$).

\section{Discussion}
In the present work, we studied the timing and spectra properties of low-luminosity highly obscured AGN NGC~4941 using combined data from {\it NuSTAR} and {\it Swift}/BAT in $3-150$~keV range {in detail}. The broad-band investigation allowed us to understand the properties of the nucleus and obscuration of the AGN in the present study.

We observed similar variability in $3-10$~keV and $10-60$~keV energy ranges with $F_{\rm var} \sim 14\%$. The observed variability is found to be comparable to other Seyfert galaxies (Hernandez et al. 2015). In general, the primary photons and reprocessed photons dominate in $3-10$~keV and $10-60$~keV energy ranges, respectively. Therefore, the variability is expected to be different in  different energy bands. However, in NGC~4941, we observe a similar contribution of the primary and reprocessed emission in $3-10$~keV and $10-60$~keV energy bands. The ratios between the primary and reprocessed flux in $3-10$~keV and $10-60$~keV energy bands are estimated to be $F_{\rm PL}/F_{\rm refl} \sim 1.01$ and $\sim 0.98$, respectively, from the slab model. This is the probable reason for similar variabilities in above energy bands.

For our study, we used the phenomenological slab model and physically motivated {\tt mytorus} model. Although both models gave us similar statistics, a clear discrepancy of the derived parameters is visible. It is clear from Figure~2 and Figure~4 that the reprocessed emission is greater in the {\tt mytorus} model compared to that in the slab model. This can be seen from the derived parameters $A_{\rm S}$ and $R_{\rm refl}$. Although $A_{\rm S}$ is not the same as $R_{\rm refl}$, it can be used as a proxy of reflection fraction. From our spectral analysis, we obtained $R_{\rm refl} < 0.89$ and $A_{\rm S} = 3.15\pm 0.32$ from the slab model and {\tt mytorus} model, respectively. The slab model indicates that the continuum emission dominates while the {\tt mytorus} model indicates the opposite. This discrepancy arises due to different treatment of the reprocessed emission by a different model. The slab model does not consider the complex structure of the torus.  Hence, this model may not calculate the reprocessing component correctly. On the other hand, {\tt mytorus} considers the complex structure of the torus. Therefore, it describes the reprocessed emission more accurately. Different treatments of the reprocessed emission affects the spectral shape, as the photon index was flatter in the {\tt mytorus} model.

The relative normalization obtained from the {\tt mytorus} model was $A_{\rm S}=3.15 \pm 0.32$. The deviation from unity indicates that either the reprocessed emission was delayed or the torus covering factor deviated from 0.5 or multiple absorbers with different column densities exist along the line of sight (Yaqoob 2012).

Zhao et al. 2021 studied NGC 4941 with {\tt borus} model (Balokovic et al. 2018) and estimated the line of sight column density and torus column density as $N_{\rm H}^{\rm los} = 9.5^{+9.5}_{-3.5} \times 10^{23}$ cm$^{-2}$ and $N_{\rm H}^{\rm tor} = 3.2^{+5.5}_{-2.5} \times 10^{24}$ cm$^{-2}$, respectively. Garcia-Burillo et al. 2021 also studied the source assuming the torus covering factor as 1 and found the line of sight column density and torus column density as $N_{\rm H}^{\rm los} = 4.2 \pm 1.1 \times 10^{23}$ cm$^{-2}$ and $N_{\rm H}^{\rm tor} = 1.4 \pm 0.3 \times 10^{24}$ cm$^{-2}$, respectively. Georgantopoulos et al. 2019 applied a different configuration of {\tt mytorus} model to the same set of {\it NuSTAR} and BAT data and estimated the equatorial column density $N_{\rm H}^{\rm Eq} = 1.2^{+0.9}_{-0.4} \times 10^{24}$ cm$^{-2}$. In their analysis, they used {\tt comptt} model (Titarchuk 1994) as the primary emission. However, we used {\tt powerlaw} as the primary emission in our analysis. We obtained the line of sight column density as $N_{\rm H}^{\rm los} = (0.76\pm 0.09) \times 10^{24}$ cm$^{-2}$. The {\it Suzaku} observation in June 2012 reported the line of sight column density as $N_{\rm H}^{\rm los} = (0.73\pm 0.19) \times 10^{24}$ cm$^{-2}$, which is consistent with our result (Kawamuro et al. 2013). From the previous observation of NGC~4941, a variable $N_{\rm H}^{\rm los}$ was observed though the timescale of variability could not be determined. In July 1996, {\it ASCA} observed the source in the CT state with $N_{\rm H}^{\rm los} = (0.99\pm0.12) \times 10^{24}$ cm$^{-2}$ (Maiolino et al. 1998). {\it BeppoSAX} observed the source six months later and found the source in Compton-thin state with $N_{\rm H}^{\rm los} = 0.45^{+0.24}_{-0.14} \times 10^{24}$ cm$^{-2}$ (Terashima et al. 2002). From the past observations, it is also clear that NGC~4941 shows variability of $N_{\rm H}$ in months timescale although the variability in shorter timescales cannot be ruled out. The spectral analysis with the {\tt mytorus} model also revealed the averaged global column density of the obscured medium as $N_{\rm H}^{\rm tor} = 3.09^{+1.68}_{-1.01} \times 10^{24}$ cm$^{-2}$ (present work).

In our study, we found a strong Fe K$_\alpha$ line with EW = $1.03^{+0.07}_{-0.13}$~keV. A strong signature of Fe K$_\alpha$ line was also observed in past. The {\it ASCA} and {\it BeppoSAX} observations reported the EW of Fe K$_\alpha$ line as $0.57\pm0.23$~keV and $1.6\pm0.8$~keV, respectively (Maiolino et al. 1998, Terashima et al. 2002). In contrast, a comparatively weak iron line with EW = $0.38\pm0.08$~keV was detected in the {\it Suzaku} data (Kawamuro et al. 2013). Kawamuro et al. 2013 found that the flux of the iron K$_\alpha$ line was consistent across the observations. The Fe K$_\alpha$ line flux was estimated to be $1.0\pm0.4$, $1.2\pm0.6$ and $0.67\pm0.12 \times 10^{-5}$ photons cm$^{-2}$ s$^{-1}$ from the {\it ASCA}, {\it BeppoSAX}, and {\it Suzaku} observations, respectively (Terashima et al. 2002, Cardmone et al. 2007, Kawamuro et al. 2013). In our analysis, the Fe K$\alpha$ line flux is estimated to be $0.76\pm 0.18 \times 10^{-6}$ photons cm$^{-2}$ s$^{-1}$ which is consistent with the previous observations. The consistent iron line flux indicated that the reflection flux originating at the torus, therefore, is constant over the years (Kawamuro et al. 2013).

We estimated the intrinsic luminosity of the source in $2-10$~keV energy range as $L_{\rm 2-10}^{\rm int} = 2.13 \times 10^{41}$ ergs s$^{-1}$. Considering the bolometric correction factor $\kappa_{\rm 2-10}=20$ (Vasedevan et al. 2009), we obtained the bolometric luminosity $L_{\rm bol} = (4.26 \pm 1.32) \times 10^{42}$ ergs s$^{-1}$. This indicates that the Eddington ratio $\lambda_{\rm Edd} = L/L_{\rm Edd} = 0.0041\pm0.0013$. The observed Eddington ratio is consistent with the other Seyfert~2 galaxies (Wu et al. 2004). The $2-10$~keV intrinsic luminosity of the source, estimated from the {\it Suzaku} observation which was carried out $\sim 3.5$ years prior to the {\it NuSTAR} observation, was $1.98\times10^{41}$ ergs s$^{-1}$. This suggests a $\sim 10\%$ change in the mass accretion rate in the source over $\sim 3.5$~years.

The X-ray emitting Compton corona is characterized by the hot electron temperature ($kT_{\rm e}$) and the optical depth ($\tau$). One can make crude approximation of $kT_{\rm e}$ and $\tau$ without detailed modelling if the photon index ($\Gamma$) and cut-off energy ($E_{\rm cut}$) are known. In general, $E_{\rm cut}=2-3~kT_{\rm e}$ (Petrucci et al. 2001). Specifically, 
\vspace{1.5em}
\\
$~~~~~~~~~~~~~~~~~~~E_{\rm cut} = 2~kT_{\rm e}$~~~~~~~~~~~~~~~~~~~~~~~~~for $\tau \leq 1$.
\\
$~~~~~~~~~~~~~~~~~~~~~~~~~ = (1+\tau)~kT_{\rm e}$~~~~~~~~~~~~~~~~for $1<\tau<2$.
\\
$~~~~~~~~~~~~~~~~~~~~~~~~~ = 3~kT_{\rm e}$~~~~~~~~~~~~~~~~~~~~~~~~~for $\tau \geq 2$.
\vspace{1.5em}
\\
From our spectral analysis with the slab model, we found that the cut-off energy of the source is $E_{\rm cut} = 177^{+92}_{-16}$~keV. From this, the hot electron temperature is estimated to be either $88^{+46}_{-8}$~keV or $59^{+31}_{-5}$~keV, or between these. One can calculate the optical depth from the following equation (Titarchuk 1994, Zdziarski et al. 1996),

\begin{equation*}
    \tau \approx \sqrt{\frac{9}{4}+\frac{m_{\rm e}c^2}{kT_{\rm e}}\frac{3}{(\Gamma -1)(\Gamma + 2)}} - \frac{3}{2}.
\end{equation*}

Using this relation, we obtained $\tau \sim 2$ (for $kT_{\rm e} \sim 88$~keV) or $\tau \sim 2.7$ (for $kT_{\rm e} \sim 59$~keV). From this, it is clear that the the hot electron temperature, $kT_{\rm e} = 59^{+46}_{-5}$~keV and the optical depth $\tau = 2.7^{+0.2}_{-1.6}$. Balokovic et al. 2020 also estimated the cutoff energy of the source as $E_{\rm cut} = 110^{+u}_{-60}$~keV. They calculated the optical depth of the corona as $\tau = 3.6$. Though the value of $\tau$ is marginally different from our estimation, the cut-off energy is consistent with our measurement.

We also tried to estimate the Compton corona temperature and the optical depth from detailed spectral modelling. We replaced the {\tt power-law} in the slab model with {\tt comptt} model. We carried out spectral fitting for spherical and slab coronae. We obtained $kT_{\rm e} = 67^{+145}_{-54}$~keV and $\tau = 4.18^{+4.28}_{-3.65}$ for the spherical corona and $kT_{\rm e} = 74^{+136}_{-59}$~keV and $\tau = 1.81^{+2.40}_{-1.26}$ for the slab corona. Although, the {\tt comptt} model yielded a higher uncertainty in the obtained values, it is consistent with the crude approximation. Georgantopoulos et al. 2019 also estimated the Compton corona temperature from the fitting with {\tt mytorus} model as $kT_{\rm e} = 47^{+216}_{-34}$~keV. Our finding is  consistent with this.

\section{Conclusion and Summary}
We studied NGC~4941 using the combined data of {\it NuSTAR} and {\it Swift}/BAT. NGC~4941 is reported to be a low luminosity AGN with $L_{\rm bol} < 10^{43}$ ergs s$^{-1}$. From the previous X-ray studies, NGC~4941 is observed to be a highly obscured AGN. Following are the findings from our work:

\begin{enumerate}
\item NGC~4941 shows similar variability in $3-10$~keV and $10-60$~keV energy ranges with $F_{\rm var} \sim 14\%$. 
\item We obtained the line of sight column density $N_{\rm H}^{\rm los} = (0.76 \pm 0.09) \times 10^{24}$ cm$^{-2}$ from the spectral analysis with {\tt mytorus} model. The slab model also showed similar value. The averaged global torus density is found to be $N_{\rm H}^{\rm tor} = 3.09^{+1.68}_{-1.01} \times 10^{24}$ cm$^{-2}$ suggesting as Compton-thick.
\item We find evidence of strong reflection in the source from both slab and {\tt mytorus} models. A deviation of relative normalization ($A_{\rm S}$) from unity indicates a delayed reprocessed emission.
\item The iron K-line emission is observed to be strong with EW = $1.03^{+0.07}_{-0.13}$~keV.
\item The bolometric luminosity is estimated to be $L_{\rm bol} = (4.26 \pm 1.32) \times 10^{42}$ ergs s$^{-1}$. We obtain the Eddington ratio of the source as $\lambda_{\rm Edd}=0.0041\pm0.0013$.
\item The cut-off energy of the source is found to be as $E_{\rm cut}=177^{+92}_{-16}$~keV. From this, we estimate the hot electron temperature and optical depth of the Compton cloud as $kT_{\rm e}=59^{+31}_{-5}$~keV and $\tau=2.7^{+0.2}_{-1.6}$.
\end{enumerate}

\vspace{2em}

\section*{Acknowledgements}
We acknowledge the anonymous reviewer for the helpful comments and suggestions which improved the paper. The research work at Physical Research Laboratory, Ahmedabad, is funded by the Department of Space, Government of India. This work has made use of the data and/or software provided by the High Energy Astrophysics Science Archive Research Center (HEASARC), which is a service of the Astrophysics Science Division at NASA/GSFC and the High Energy Astrophysics Division of the Smithsonian Astrophysical Observatory. This research has made use of the {\it NuSTAR} Data Analysis Software ({\tt NuSTARDAS}) jointly developed by the ASI Space Science Data Center (SSDC, Italy) and the California Institute of Technology (Caltech, USA).

\appendix

\vspace{1em}


\end{document}